\documentclass[preprint,aps,prb,onecolumn,superscriptaddress,floatfix,longbibliography,11pt]{revtex4-2}

\usepackage{amsmath,amssymb} % math symbols
\usepackage{bm} % bold math font
\usepackage{graphicx} % for figures
\usepackage{textcomp} % This package is just to give the text quote '
\usepackage[nolist,nohyperlinks]{acronym} % acronym management
\usepackage{float}
\usepackage{upgreek}
\usepackage{mdframed}
\usepackage{xcolor}
\usepackage{setspace}
\usepackage{siunitx} % SI unit formatting
\usepackage{xr} % external document references
\usepackage{enumitem} % custom list formatting
\usepackage{hyperref} % hyperlinks
\hypersetup{
    colorlinks=true,
    urlcolor= blue,
    citecolor=blue,
    linkcolor= blue,
}
\newcommand{\mytitle}{High-power quantum walk frequency combs}
\begin{document}

\title{\mytitle}

\author{Theodore P. Letsou}
\email{tletsou@g.harvard.edu}
\affiliation{\textbf{These authors contributed equally to this work.}}
\affiliation{Harvard John A. Paulson School of Engineering and Applied Sciences, Harvard University, Cambridge, MA 02138, USA}
\affiliation{Department of Electrical Engineering and Computer Science, Massachusetts Institute of Technology, Cambridge, MA 02142, USA}

\author{Johannes Fuchsberger}
\affiliation{\textbf{These authors contributed equally to this work.}}
\affiliation{Institute of Solid State Electronics, TU Wien, 1040 Vienna, Austria}

\author{Nikola Opa{\v{c}}ak}
% \author{Nikola Opacak}
\affiliation{Institute of Solid State Electronics, TU Wien, 1040 Vienna, Austria}
\affiliation{Harvard John A. Paulson School of Engineering and Applied Sciences, Harvard University, Cambridge, MA 02138, USA}

\author{Dmitry Kazakov}
\thanks{\textit{Now at imec}}
\affiliation{Harvard John A. Paulson School of Engineering and Applied Sciences, Harvard University, Cambridge, MA 02138, USA}

\author{Benedikt Schwarz}
\email{benedikt.schwarz@tuwien.ac.at}
\affiliation{\textbf{These authors contributed equally to this work.}}
\affiliation{Institute of Solid State Electronics, TU Wien, 1040 Vienna, Austria}
\affiliation{Harvard John A. Paulson School of Engineering and Applied Sciences, Harvard University, Cambridge, MA 02138, USA}

\author{Federico Capasso}
\email{capasso@seas.harvard.edu}
\affiliation{\textbf{These authors contributed equally to this work.}}
\affiliation{Harvard John A. Paulson School of Engineering and Applied Sciences, Harvard University, Cambridge, MA 02138, USA}

\date{\today}

\begin{abstract}
%mention approx 1mW per line and wavelength simulatnously for comb comunity
\textbf{
High-power, broadband frequency combs generated by semiconductor lasers have profound implications for on-chip spectroscopy.
Here, we present a dry-etched racetrack quantum cascade laser that uses resonant radio-frequency injection to produce a unidirectional “quantum walk” frequency comb at mid-infrared wavelengths.
Efficient light outcoupling from the racetrack resonator provides more than 100\,mW of continuous-wave output power at room temperature, with beam quality on par with that of Fabry--Perot lasers.
Experimental waveform reconstruction confirms that the combs are frequency-modulated, rather than amplitude-modulated (as in active mode-locking).
We show excellent agreement between the experimental waveforms and numerical simulations, which are based on a reduced Maxwell--Bloch model of the laser and include fast gain dynamics, finite group velocity dispersion, and large Kerr nonlinearity.
Furthermore, our optimized device architecture --- featuring thick Si$_3$N$_4$ passivation and reduced parasitic capacitance --- enables modulation bandwidths exceeding 10\,GHz.
Combined with high output powers and the potential for monolithic integration of multiple ring lasers and waveguide couplers, these advances pave the way to fully integrated dual-comb spectrometers.}

\end{abstract}

\maketitle
\section{\label{sec:Intro}Introduction}

Frequency combs emerge when optical lines synchronize through a balance of optical gain and loss, nonlinearity, and dispersion, resulting in a broadband spectral output with each line separated by a common difference frequency \cite{Fortier2019}.
They can form in various systems with center wavelengths ranging from the UV to the THz, making them ideal sources for spectroscopy \cite{Picqu2019, Hermans2022}.
Among numerous potential platforms, semiconductor lasers stand out as prominent frequency comb sources due to their compact footprint, operational versatility, and relatively broad spectral coverage \cite{Dong2020, Sterczewski2022, Sterczewski2020}.
In particular, semiconductor lasers with short gain recovery time, such as quantum cascade lasers (QCLs), are especially promising because they can self-mode-lock when operated under direct current (DC) bias. \cite{Hugi2012, Faist2016, Dong2023, Hillbrand2020}.
Unlike traditional active mode-locking, which relies on strong radio-frequency (RF) injection to create a short temporal net gain window for pulse formation, self-mode-locking produces a frequency-modulated (FM) comb, resulting in a chirp of the instantaneous frequency of the laser, rather than pulse emission \cite{Singleton2018, Opaak2019}.
This behavior is partly attributed to the picosecond gain recovery times, $T_1$, of QCLs, their exceptionally high Kerr nonlinearities, and the coupling between forward and backward propagating waves in the laser cavity \cite{Piccardo2019, Opaak2021}.

Recently, new regimes of frequency comb formation have been discovered in ring QCLs.
Unlike Fabry--Perot cavities, ring cavities support unidirectional operation: the wave circulates clockwise (CW) or counterclockwise (CCW) around the cavity in steady state.
As a result, spatial hole burning does not induce the gain competition that would destabilize the single-mode solution, as it does in Fabry--Perot lasers \cite{Mansuripur2016}.
Nonetheless, single-mode instabilities can still arise spontaneously under DC bias.
Solitons and turbulent frequency combs—both experimentally observed in ring QCLs—are well described by the Ginzburg--Landau formalism \cite{Piccardo2020, Opacak2024, Meng2020, Meng2021, Micheletti2023, Jaidl2021}, highlighting how ring lasers can exhibit complex comb dynamics distinct from those found in traditional Fabry--Perot geometries.

Another method to destabilize the single-mode solution is to modulate the laser current at a frequency resonant with the cavity roundtrip frequency \cite{Opaak2024_2, Schneider2021}, in many ways similarly to electro-optic frequency combs in materials with non-zero Pockels coefficient~\cite{Hu2022, Zhang2019_2}.
This method leads to a ``quantum walk comb,” so named for the comb formation can be pictured as a random walk of photons across the lattice sites in a synthetic frequency space \cite{Heckelmann2023}. 
In the frequency domain, the quantum walk comb manifests itself as broadband frequency comb with a high-order Hermite-Gaussian spectral envelope, unlike spectra with Gaussian envelopes in mode-locked lasers with slow gain recovery~\cite{Haus1975}.
The RF injection serves to lock the modes together, give precise control over the laser's repetition rate \cite{https://doi.org/10.48550/arxiv.2411.11210}, and stabilize the comb's output, making them ideal sources for dual-comb spectrometers \cite{Consolino2020, Villares2014}. 
However, one requirement for the emergence of this state is unidirectional laser operation.
This necessitates extra care during device fabrication to ensure low sidewall roughness on the deeply etched waveguide, and requires low-back-reflection outcouplers that are not routinely available in the standard QCL process.
Therefore, most demonstrations of ring QCLs utilize waveguide bending losses to achieve light outcoupling, which limits their output powers to submilliwatt levels, even though their intracavity powers may reach hundreds of milliwatts.
For example, surface emitting ring QCLs with gratings routinely reach sub-watt output powers at room temperature \cite{Bai2011}.

In this work, we present a ring QCL in a racetrack (RT) geometry with a directional coupler capable of emitting a broadband frequency comb via resonant RF injection at the laser's round-trip frequency.
The waveguide-based coupler with cleaved facets enables output powers exceeding 100 mW with excellent beam quality. 
Optimized modulation sections with thick Si$_3$N$_4$ passivation allow high-frequency operation ($f_{\text{3 dB}} > 10$ GHz) by reducing the parasitic capacitance of the laser contact. 
Our devices are fabricated with a single dry-etch without the need for epitaxial regrowth or regrown passive waveguides \cite{Kazakov2024, Wang2022}.
Furthermore, our experimental results are supported by a theoretical model based on a laser master equation including resonant gain modulation derived from the Maxwell-Bloch equations.
Numerical simulations are in excellent agreement with experimental waveform reconstruction measurements and further identify the Kerr nonlinearity as a dominant factor in the formation of the frequency comb.
We also identify regions of instability at high levels of RF injection, where the laser spectra deviate from Hermite-Gaussian envelopes, offering an insight on the bandwidth limitations of the quantum-walk comb. 
Our device platform is robust, well-behaved, and offers a highly versatile solution for generating stable, broadband frequency combs on a chip.

\section{Racetrack Characterization}

\begin{figure*}[t]
    \centering
    \includegraphics[clip=true,width=\textwidth]{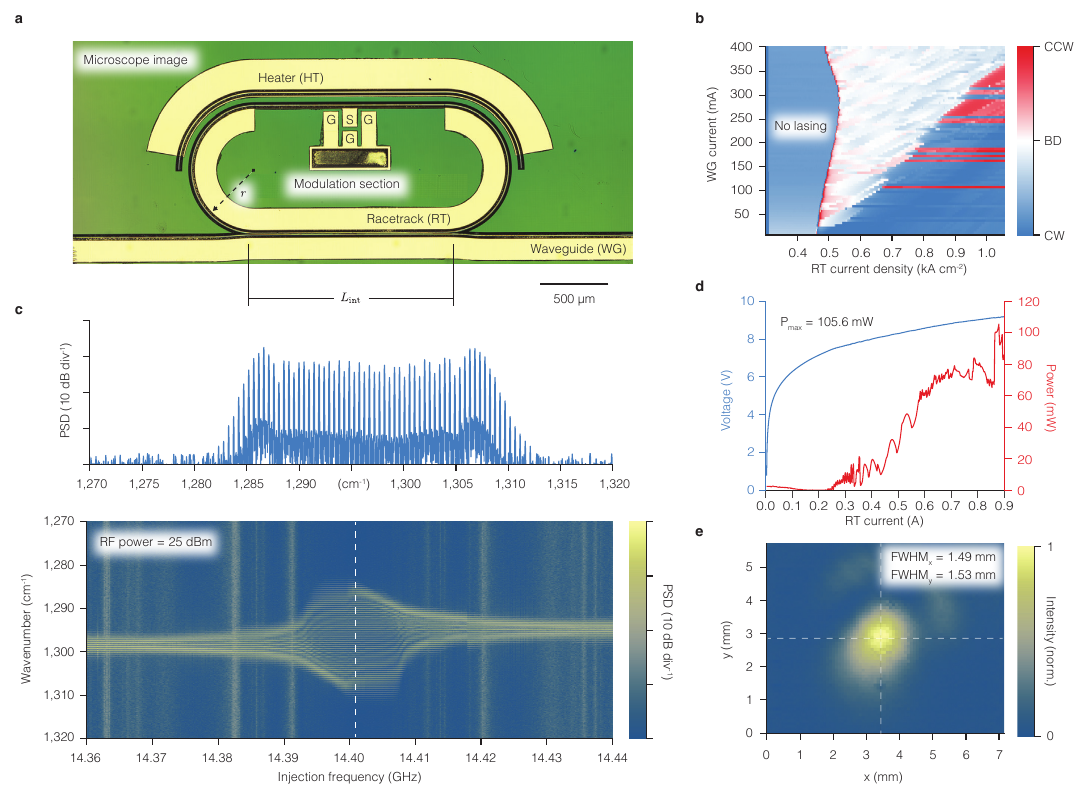}
    \caption{\textbf{Characterizing the racetrack laser.}
    \textbf{a} Microscope image of the racetrack (RT) QCL, highlighting the separate gold contacts for the RT, waveguide (WG), and integrated heater (HT). The modulation section along the straight segment of the RT is designed for ground-signal-ground (GSG) probing with a thick Si$_3$N$_4$ insulation layer to minimize parasitic capacitance.
    \textbf{b} Measured output from each WG facet as a function of RT and WG bias, showing a transition from bidirectional (BD) to counterclockwise (CCW) or clockwise (CW) unidirectional operation above threshold.
    \textbf{c} Power spectral density (PSD) of the RT QCL under a 25\,dBm RF drive near its 14.4\,GHz round-trip frequency, displaying a broad Hermite-Gaussian envelope spanning over 30\,cm$^{-1}$.  This spectra is a slice taken from a spectral heat map which shows the spectrum of the quantum walk RT as a function of RT injection frequency.
    \textbf{d} Light-current-voltage (LIV) plot from a single WG facet at room temperature (epi-up mounting), reaching a maximum output of 105.6\,mW. 
    \textbf{e} Far-field beam profile measured approximately 1\,m away from the WG facet, showing tight collimation with a near-circular beam shape.}
    \label{fig:1}
\end{figure*}

Figure~\ref{fig:1}\textbf{a} shows a microscope image of the dry-etched device.
The top laser contact consists of several gold sections: one for biasing the racetrack (RT), one for the waveguide (WG), and one for the integrated heater (HT).
The bent section of the RT has a radius of $r=500$ \textmu m. 
The straight section of the RT that is evanescently coupled to the WG has a length of $L_{\text{int}} = 1.5$ mm.
Similar laser geometries have been shown to emit a plethora of coherent laser states, such as homoclons, bright and dark solitons, and soliton pulse pairs \cite{Piccardo2020,Opaak2024,soliton, Letsou2025}.
In addition to the DC bias sections, the device incorporates a high-frequency modulation section along the straight segment of the RT.
To minimize parasitic capacitance, and increase the modulation efficiency, we reduced the gold contact area over the modulation section and deposited a thick (1,500 nm) Si$_3$N$_4$ insulation layer.
Planarized ground contacts enable the use of standard ground-signal-ground (GSG) probes, allowing efficient modulation of the device during operation.

The RT laser undergoes spontaneous symmetry breaking when biased above its laser threshold with DC ($J_{\text{th}} \sim 0.46 \text{ kA cm}^{-2}$), converting its intracavity field from bidirectional to unidirectional. 
This transition can be seen when measuring the laser output from both WG facets while sweeping the bias on both the RT and WG sections, as shown in Fig.~\ref{fig:1}\textbf{b}.
The ratio of the signals from the balanced pair of detectors allows us to determine the proportion of light circulating in either the CW or CCW direction.
We observe that the symmetry-breaking point occurs at higher RT currents when the WG is biased.
Once the intracavity field is unidirectional, we apply a 25 dBm RF signal around the repetition frequency of our RT, which is $\sim 14.4\text{ GHz}$.
On resonance, the single-mode output of the laser spectrally broadens to a frequency comb with a Hermite-Gaussian envelope spanning over 30 cm$^{-1}$, as confirmed by measurements of the laser's power spectral density (PSD) shown in Fig.~\ref{fig:1}\textbf{c}.
Our laser output exceeds 100 mW from a signal WG facet at room temperature when mounted epi-up on a copper submount (Fig.~\ref{fig:1}\textbf{d}).
The output is also tightly collimated, with a small beam diameter; its full width at half maximum (FWHM) is approximately 1.5 mm when measured one meter from the facet (Fig.~\ref{fig:1}\textbf{e}).
These output characteristics are typically only achievable in narrow waveguide edge-emitting Fabry-Perot-type geometries, but our outcoupling scheme allows us to efficiently extract light from the RT without perturbing the unidirectional physics required to achieve resonant spectral broadening.

\section{Interpreting the spectral broadening}

The physical effects responsible for the spectrally broadened frequency comb output are revealed by examining the unidirectional master equation for lasers with fast gain dynamics. 
The equation shown below is derived from the full set of Maxwell--Bloch equations under an appropriate set of approximations valid for QCLs, as indicated by numerous examples of agreement between experimental waveforms and those predicted by the master equation \cite{Opaak2019}:
\begin{equation}
\label{eq:master_equation}
\bigl( n_c \,\partial_t + \partial_z \bigr) E
= \frac{g(P)}{2} \,(1 + i \alpha)
\bigl[\, E \;-\; T_2\,\partial_t E \;+\; T_2^2\,\partial_t^2 E \bigr]
\;+\; i\,\frac{k''}{2}\,\partial_t^2 E
\;+\; i\,\beta\,\lvert E \rvert^2 E
\;-\;\frac{\alpha_w}{2}\,E,
\end{equation}
where $E$ is the slowly varying electric field envelope, $P=\lvert E\rvert^2$ is the optical intensity, $g(P)$ is the saturated gain, $T_2$ is the decoherence time, $\alpha$ is the linewidth enhancement factor (LEF), $k''$ is the group velocity dispersion, $\beta$ is the Kerr nonlinearity, and $\alpha_w$ accounts for losses.
The current in the modulation section is modeled by a sinusoidal modulation added to the DC current, $J_{\mathrm{DC}} + J_{\mathrm{mod}}\sin(\omega_{\mathrm{mod}} t)$, which modulates the gain through $g(P) \sim J$ \cite{Hillbrand2020_2}.
However, a modulation in the gain must introduce a phase modulation in the field to create resonant coupling between each of the cavity modes.
Therefore, only terms that phase shift the electric field ($\alpha$ or $\beta$) can be responsible for frequency-comb formation.
Through numerical simulations, we found that a nonzero $\beta$ is a dominant factor leading to spectral broadening via current modulation.

\begin{figure*}[t]
    \centering
    \includegraphics[clip=true,width=\textwidth]{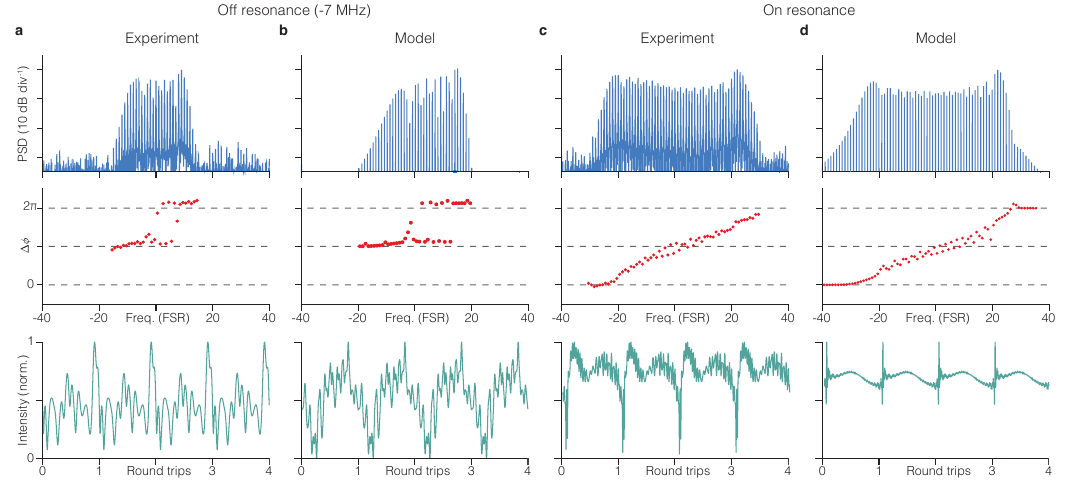}
    \caption{\textbf{Experiment and simulation of the quantum‐walk racetrack.}
    Each column shows three plots (top to bottom): the PSD of the laser states, the intermodal phase differences, and the waveform intensity.
    \textbf{a},\,\textbf{b}~Experimental reconstruction and simulation, respectively, of a racetrack operated 7\,MHz detuned from resonance.
    In this off‐resonance configuration, the phases exhibit \(\pi\) phase jumps reminiscent of a Bessel‐type spectrum, along with a slow per‐round‐trip modulation.
    \textbf{c},\,\textbf{d}~Experimental reconstruction and simulation for a racetrack operated on resonance.
    Here, the phase differences span from \(0\) to \(2\pi\) every round‐trip, yielding a quasi‐constant waveform that features a short spike at the turnaround point.}
    \label{fig:2}
\end{figure*}

We perform SWIFTS to experimentally examine the phase dynamics and measure the waveform of the RT QCL under 21 dBm of RF injection \cite{Han2020, Burghoff2015}.
When the RF injection is far ($\sim 7\,\text{MHz}$) detuned from resonance, we observe only modest spectral broadening (Fig.~\ref{fig:2}\textbf{a}), spanning approximately 30 optical lines (plotted in normalized frequency units, where one unit the free-spectral range of the laser).
The intermodal phase differences, which exhibit $\pi$ phase jumps across the center of the spectrum, strongly resemble a Bessel-type phase modulation.
As a result, the reconstructed waveform—calculated from both the PSD and intermodal phase differences—shows a low-contrast intensity modulation during each round-trip.
Simulations based on the master equation, shown in Fig.~\ref{fig:2}\textbf{b}, are in excellent agreement with our measurements, confirming the formation of Bessel-type phases.
The plotted waveform is taken after simulated 10{,}000 round-trips, with $\beta = 50 \times 10^{-12} \ \mathrm{V\ m^{-1}},\ k^{''}=0 \text{ fs}^{2}\ \text{mm}^{-1}$ and $\alpha = 0$.
(See the Supplementary Information for additional details on the simulations.)
When the RF injection is tuned on-resonance with the round-trip frequency of the RT QCL, we measure a comb containing more than 60 optical lines, as shown in the top panel of Fig.~\ref{fig:2}\textbf{c}.
SWIFTS reveals a splayed-phase intermodal phase configuration spanning from $0$ to $2\pi$, a profile often observed in free-running Fabry–Perot semiconductor lasers with bidirectional field propagation.
Our experimental results agree well with the master equation simulation shown in Fig.~\ref{fig:2}\textbf{d}, now with a current modulation resonant with the round-trip frequency.

\section{RF optimization on a dry-etched platform}

\begin{figure*}[t]
    \centering
    \includegraphics[clip=true,width=\textwidth]{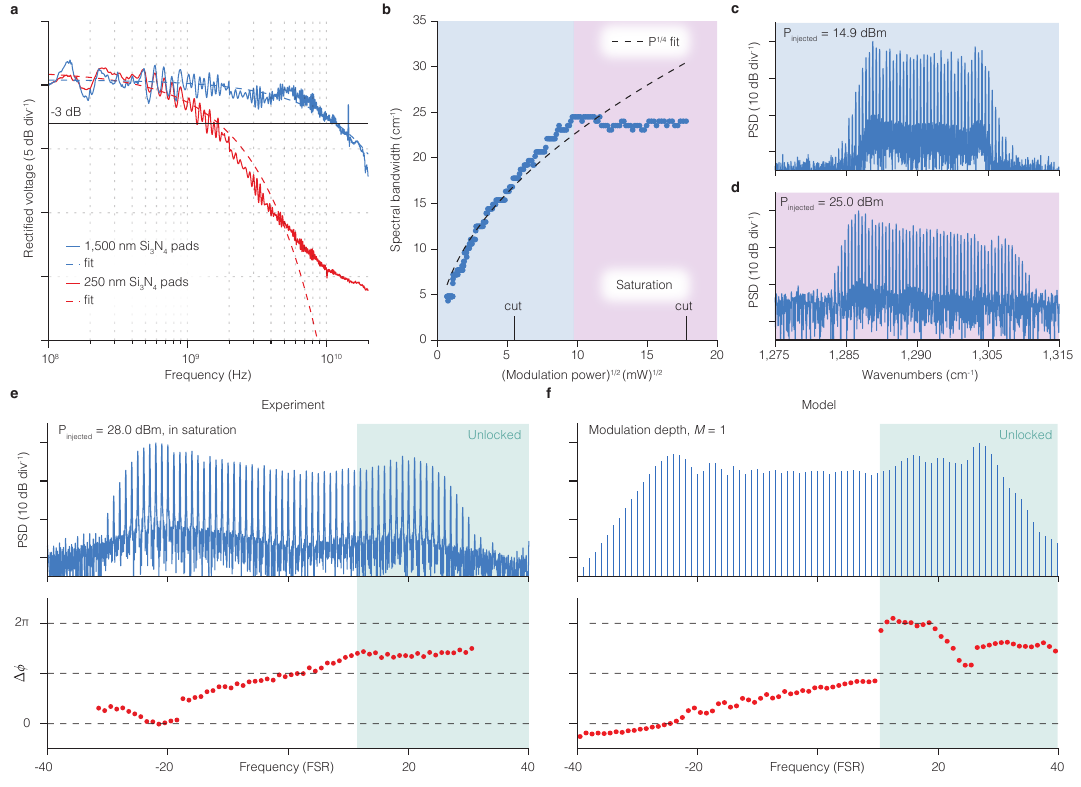}
    \caption{\textbf{RF optimization and comb power scaling.}
    \textbf{a} Rectified voltage vs.\ modulation frequency for two RT QCL designs with different Si$_3$N$_4$ insulation thicknesses (250\,nm vs.\ 1{,}500\,nm). Increasing the Si$_3$N$_4$ thickness reduces parasitic capacitance, raising the 3\,dB cutoff frequency from about 1.5\,GHz to above 10\,GHz (fits shown as dashed lines).
    \textbf{b} Measured QCL spectral bandwidth (in cm$^{-1}$) as a function of the square root of the modulation power, $P^{1/2}$. The data follow the expected $P^{1/4}$ scaling (dashed line) until the comb saturates at $P\sim100$\,mW.
    \textbf{c} Representative RF-driven comb spectrum at moderate injected power ($14.9$\,dBm), showing a high dynamic range and broad coverage.
    \textbf{d} At higher RF injection ($25.0$\,dBm), the spectrum destabilizes into an incoherent multimode output with a raised noise floor, indicating the onset of saturation at high drive levels.
    \textbf{e} Experimental SWIFTS reconstruction under high injection power ($28.0$\,dBm) reveals a partially unlocked comb, as indicated by the flattening of the reconstructed phase differences (green box).
    \textbf{f} Simulation of an on-resonance modulation with high modulation depth ($M=1$) is in close agreement with the measurement. A similar flattening of the phases near the edge of the spectrum, deviating from the linear chirp, shows the unlocking of the comb.}
    \label{fig:3}
\end{figure*}

The injection frequency required for spectral broadening depends on the device geometry---as the device size is reduced, its repetition rate increases.
A typical compact ring laser may have a repetition rate of 10 to 20 GHz.
Therefore, reducing parasitic capacitance is imperative for increasing the modulation bandwidth and achieving a broad spectral output.
Wet-etched QCLs often have epitaxially regrown insulation layers, such as Fe:InP.
These layers provide electrical isolation, enhance the device's thermal performance, and, importantly, reduce parasitic capacitance between the laser contact and ground.
Dry-etched ridge lasers are often incompatible with epitaxial regrowth due to their rough etch profiles; therefore, thin layers of Si$_3$N$_4$ are deposited along the etched sidewalls to isolate the laser contact from ground.
The thin insulation layer leads to a relatively low RF cutoff, as indicated by the rectification measurement for an RT QCL with a 250 nm thick Si$_3$N$_4$ insulation layer, as shown in Fig.~\ref{fig:3}\textbf{a}.
To decrease the parasitic capacitance, we deposit a 1,500 nm Si$_3$N$_4$ layer on one side of the modulation section to bridge the top contact and ground.
This minor change---in addition to reducing the top contact area of the modulation section---greatly improves the lasers RF performance, increasing the 3 dB cutoff frequency from $\sim 1.5$ GHz to $>10$ GHz (Fig.~\ref{fig:3}\textbf{a} and Supplementary Information).

Unlike actively mode-locked lasers, whose optical bandwidth follows an $M^{1/4}$ relation, where $M$ is the modulation depth \cite{Kuizenga1970}, the quantum walk comb follows an $M^{1/2}$ relation \cite{Heckelmann2023}.
The modulation power $P$ injected into the laser is related to the modulation depth $M$ by $P \propto M^2$.
Therefore, we expect the comb bandwidth to scale as $P^{1/4}$.
This relation is confirmed by sweeping the strength of the injected RF power shown in Fig.~\ref{fig:3}\textbf{b} until $P=100$ mW (20 dBm, other devices have differing saturation points).
After this power level the laser's spectral bandwidth saturates, transforming from the high-dynamic-range comb shown in Fig.~\ref{fig:3}\textbf{c} into an incoherent, multimode spectrum whose noise floor increases by 10 dB (Fig.~\ref{fig:3}\textbf{d}).
To illustrate the unlocking of the comb state, we also performed SWIFTS measurements in the saturation regime, as shown in Fig.~\ref{fig:3}\textbf{e}.
The intermodal phases exhibit flattening at the edges of the spectrum, while the central modes retain a linear chirp.
The numerical simulations emulating the high-power RF injection by imposing a larger modulation depth ($M=1$) confirm this observation  (Fig.~\ref{fig:3}\textbf{f}).
The simulated decoherence of the intermodal phases on the blue side of the laser spectrum aligns well with our measurements.
The simulations signify that comb destabilization originates from the spectral edges, where insufficient gain prevents coherent lasing: in essence, the stable FM comb bandwidth surpasses the laser’s gain bandwidth.  
The modes on the edge of the spectrum are not intense enough to saturate the gain and prevent other modes from lasing, thus chaotic multimode emission ensues, causing the comb to unlock \cite{Opaak2019, Burghoff2020}.

\section{\label{sec:outlook} Outlook}

We introduced a dry-etched laser platform to generate stable frequency combs based on resonant RF injection, employing high-frequency modulation sections fabricated alongside the laser.
These frequency combs arise from the QCL's fast gain dynamics and large intrinsic Kerr nonlinearity, stemming from intersubband transitions, which is orders of magnitude higher than those in bulk crystal materials.
The reconstructed spectral phases and waveforms closely match simulations based on the reduced Maxwell--Bloch model for a unidirectional ring cavity.
The waveguide directional coupler efficiently extracts light from the racetrack through evanescent coupling, enabling output powers above 100\,mW at a center wavelength of 7.9\,\textmu m under room-temperature operation.
Advanced thermal management methods, including epi-down mounting, can further increase output powers.

Spectroscopy techniques such as dual-comb spectroscopy can fully exploit these device architectures.
The frequency comb output from nominally identical racetrack devices can be offset at the megahertz level using RF injection, allowing rapid acquisition of dual-comb absorption features.
Furthermore, multiple racetracks can be fabricated on the same chip and coupled via a shared bus waveguide and microwave transmission lines.
Combining RF and photonic engineering on the same chip will further increase modulation efficiencies, therefore reducing the required RF power to generate broadband frequency combs, as was recently demonstrated in lithium tantalate electro-optic combs \cite{Zhang2025}.
The convergence of high-speed electronics and ultrafast optical processes enables fully integrated on-chip dual-comb spectroscopy.

Beyond spectroscopic applications, these devices may also serve as useful sources for nonlinear broadening.
The linear intermode phase can be flattened by a dispersion compensator either on-chip or off-chip, leading to high peak power optical pulses~\cite{Yu2022, Tschler2021}.
Finally, recent studies indicate that the ultrafast nature of the QCL gain medium may be broadly characteristic of standard quantum well diode lasers with interband gain~\cite{https://doi.org/10.48550/arxiv.2411.08280}.
Hence, the techniques presented here could, in principle, be adapted to shorter wavelengths, opening new frontiers in semiconductor frequency combs.

\begin{acknowledgments}
T. P. Letsou thanks the Department of Defense (DoD) through the National Defense Science and Engineering Graduate (NDSEG) Fellowship Program.
This material is based on work supported by the National Science Foundation under Grant No. ECCS-2221715.
J. Fuchsberger thanks the Marshall Plan Foundation Fellowship Program.
J. Fuchsberger, N. Opa{\v{c}}ak, and B. Schwarz received funding from the European Research Council (ERC) under the European Union’s Horizon 2020 research and innovation program (Grant agreement No. 853014).
The authors gratefully acknowledge the Center for Micro- and Nanostructures (ZMNS) of TU Wien for providing the cleanroom facilities.
\end{acknowledgments}

\newpage
\bibliography{qwalk}

\end{document}